\documentclass[prl,aps,a4,epsf,8pt,twocolumn,showpacs]{revtex4}

\usepackage{amsmath}
\usepackage{amssymb}
\usepackage{graphicx}
\usepackage[usenames]{color}

\begin{document}

\newcommand{\blue}[1]{\textcolor{blue}{#1}}

\newcommand{\beq}{\begin{equation}}
\newcommand{\eeq}{\end{equation}}
\newcommand{\beqa}{\begin{eqnarray}}
\newcommand{\eeqa}{\end{eqnarray}}
\newcommand{\bmat}{\begin{displaymath}}
\newcommand{\emat}{\end{displaymath}}

\newcommand{\eq}[1]{Eq.~(\ref{#1})}

\newcommand{\lan}{\langle}
\newcommand{\ran}{\rangle}

\newcommand{\tav}[1]{\left\lan #1 \right\ran}

\title{Emergence of rigidity at the structural glass transition: a first principle computation}

\author{Hajime Yoshino$^{1}$, Marc M\'{e}zard$^{2}$}

\affiliation{$^1$Department of Earth and Space Science, Faculty of Science,
 Osaka University, Toyonaka 560-0043, Japan\\
$^2$Laboratoire de Physique Th\'eorique et Mod\`eles
  Statistiques, CNRS -  Universit\'e Paris Sud, B\^at. 100, 91405 Orsay Cedex, France}

\begin{abstract}
We compute the shear modulus
 of structural glasses from a first principle approach based on the
 cloned liquid theory. We find that the intra-state
 shear-modulus, which corresponds to the plateau modulus measured in linear
 visco-elastic measurements, strongly depends on temperature and
 vanishes continuously when the temperature is increased beyond the
 glass temperature. 
\end{abstract}


\date{\today}
\maketitle

The shear-modulus is an unambiguous measure of the mechanical
stability of materials. When one cools a glass-former below its glass
transition, there  appears a non-zero shear modulus on 
laboratory-accessible  time scales.
The understanding of the mechanism through which this rigidity emerges
at the glass transition is a basic problem in condensed matter physics.

A standard view on glasses is to regard them as very slow liquids with
extremely high shear-viscosity \cite{Angell}. Visco-elastic
measurements  show that super-cooled liquids and various soft-glassy
materials behave as solids: The elastic modulus develops at low
frequencies a  plateau, which extends to lower and lower frequencies
by lowering temperature or increasing density (see \cite{dyre-group,Weitz97} and
references therein). 
Thus glasses aquire rigidity progressively. This feature is
remarkably different from ordinary transitions from liquid to
crystal where the rigidity appears abruptly at the 1st order phase transition.

Among various theoretical attempts, the so-called random first order
theory (RFOT)\cite{RFOT} provides a useful working ground 
to study the super-cooled liquids and glasses in a unified
manner \cite{Mosaic}. 
At the mean field level it is backed up by some microscopic approaches.
On the one hand, it is intimately related to the mode-coupling theory (MCT)
concerning the dynamics at relatively high temperatures \cite{MCT}.
On the other hand, the so-called cloned liquid approach which combines
the traditional liquid theory and the replica method allows one to
compute thermodynamic static quantities at lower
temperatures\cite{mezard-parisi-1999,coluzzi-mezard-parisi-verrocchio-1999,parisi-zamponi}. 
This approach is currently the main first-principle approach to
studying properties of the glass phase. It has been so far limited to
computing thermodynamic properties, in particular the `complexity'
giving the entropy associated with the number of glass states. One of the major present
challenges 
is to understand how the nucleation processes allowing to jump between
glass states, which are not taken
into account in the simplest RFOT scenario, can be included into this
scheme. These processes are in particular crucial to explain why the
mean field prediction of a dynamical transition at the dynamical (MCT) temperature $T_c$
breaks down, and is replaced by a rapid increase of the relaxation
time when the temperature gets close to $T_c$.  Attempts in this
direction include the mosaic theory of \cite{Mosaic} and the study of
long-range interactions \cite{Franz_Kac}.

We shall extend the cloned liquid approach in order to
compute the static shear-response of glasses.
We identify the plateau modulus
mentioned above by distinguishing intra-state and inter-state
stress fluctuations. 
When applied to mesoscopic samples, this approach predicts that 
the stress vs strain curve should have  an
intermittent behaviour.

\begin{figure}[h]
\includegraphics[width=0.45\textwidth]{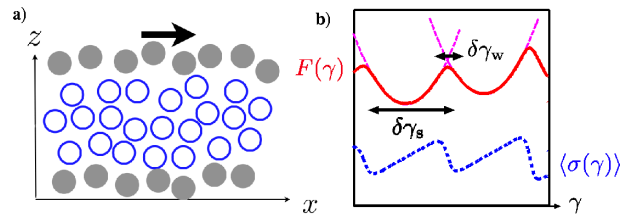}
\caption{Schematic picture of the system under static shear. 
a) The mobile particles (open circles)
are bounded by ``random walls'' (filled circle).
b) Schematic mean-field picture of energy landscape 
and stress-strain curve}
\label{fig-walls-stress-strain-rem}
\end{figure}

{\bf Models -} We consider a system of $N$ particles ($i=1,2,\ldots,N$) 
at position ${\bf r}_{i}=(x_{i},y_{i},z_{i})$ in the laboratory frame,
which are interacting with each other via two body potentials,
$
H(\{{\bf r}_{ij}\})=\sum_{i < j} v_{ij}(r_{ij}) \ ,
$
where ${\bf r}_{ij}={\bf r}_{i}-{\bf r}_{j}$ and $r_{ij}=|{\bf
  r}_{ij}|$. 
In order to study the rigidity  against simple shear deformation,
we consider a system  of particles in a container with two boundary walls which are normal 
to the $z$-axis and separated from each other by distance $L_{z}$ as shown in Fig.\ref{fig-walls-stress-strain-rem}.
To impose a shear strain $\gamma$ on the system, we simply displace 
the top wall by an amount $\gamma L_{z}$ into $x$-direction.
Then it is convenient to introduce a {\it sheared frame} with $x'$,$y'$ and $z'$
which are related to the laboratory frame as, $(x,y,z)=(x'+z' \gamma, y',z')$. 
The volume of the system $V$ and the number density
$\rho=N/V$ remain constant under this shear.

{\bf Shear-modulus - }  The total free-energy $F(\gamma)$ of the system can be written as
\beq
-\beta F(\gamma)= \ln \int_{{\cal V}}  \frac{\prod_{i}
d^{3}r'_{i}}{V^{N}} e^{-\beta H(\{r_{ij}(\gamma)\})},
\label{eq-def-free-energy}
\eeq
with $\beta^{-1}=k_{\rm B}T$ being the inverse temperature. 
The ideal gas part of the free-energy  is omitted because it does
not change.
 Note that the integration over the interior volume of the container
is taken using the sheared frame ${\bf r}'=(x',y',z')$ for which the integration range ${\cal V}$ 
is independent of the shear $\gamma$.

Taking {\it infinitesimal shear}, the free-energy can be expanded
formally as $F(\gamma)=F(0)+N\langle \sigma \rangle \gamma+\frac{N}{2}\mu\gamma^{2}+O(\gamma^{3})$,
where the shear-stress $\sigma$ and the shear-modulus $\mu$ are:
\beq
\mu= d \langle \sigma \rangle/d \gamma=\langle b \rangle -N\beta \left[ \langle \sigma^{2} \rangle -  
\langle \sigma \rangle^{2}\right] \ ,
\label{eq-def-shearmodulus}
\eeq
with
\begin{eqnarray}
&& N\sigma= \frac{d
H}{d\gamma}=
\sum_{i < j}    \left.r v'(r) \right |_{r=r_{ij}}
\hat{x}_{ij}\hat{z}_{ij} \nonumber \\
&&  N b=\frac{d^{2} H}{d \gamma^{2}}=\sum_{i <
j}\hat{z}_{ij}^{2} \left [
 r^{2}v''(r)
\hat{x}_{ij}^{2}
+  r v'(r)
(1- \hat{x}_{ij}^{2}) \right]_{r=r_{ij}} 
\label{eq-def-sigma-born}
\end{eqnarray}
where $\langle \ldots \rangle$ denotes a thermal average evaluated with
zero strain $\gamma=0$. We have introduced short hand notations like
$\hat{x}_{ij}=(x_{i}-x_{j})/r_{ij}$, and the prime stands for differentiation.
The 1st term $\langle b \rangle$ in the definition of $\mu$ in \eq{eq-def-shearmodulus} 
is called the Born term. It represents the instantaneous response of the
system against shear $G(\tau=0)$ (see below), which is finite even in
liquids. The 2nd term is the correction term due to thermal fluctuations
of the shear-stress. 

This type of fluctuation formula for the static elastic constants is
well known \cite{squire,quasi-static-zero-temperature}.
 In liquids, it is equivalent to the static limit of the Green-Kubo
 formula which relates the dynamic linear 
response against shear $\delta
\sigma(t)=\int_{-\infty}^{t}dt'G(t-t')\dot{\gamma}(t')$ to the 
shear-stress autocorrelation function $\langle
\sigma(t)\sigma(t')\rangle$ by $G(\tau)=\beta \langle
\sigma(\tau)\sigma(0)\rangle$. Linear visco-elastic measurements give
access to the complex dynamical modulus $G(\omega)=\omega
G^{*}(\omega)$. 

{\bf Boundary condition to shear - } Let us pause here to discuss more explicitely 
the boundary condition. First of all, it is obvious that the boundary
walls should {\it not} be {\it strictly} translationally invariant 
to exert shear on the system. On the other hand, we wish
the system to maintain translational invariance at least on {\it
macroscopic} scales. We thus assume that the walls
are built from a quenched {\it
random} configuration of particles, as shown in Fig.~\ref{fig-walls-stress-strain-rem} so that the system keeps
translational invariance in a {\it statistical} sense.

Because of the translational invariance at macroscopic scales, the
thermodynamic free-energy density $\lim_{V \to \infty} F(\gamma)/V$
is independent of $\gamma$. This also happens in crystals, as
discussed recently
\cite{biroli_kurchan}.
However the definition of the shear modulus (and of the solid state) is
through the linear response to a shear: the shear modulus in solids is
non-zero because of
the non-commutation of the small shear limit $\delta \gamma \to 0$ and thermodynamic limit
$V \to \infty$. The same phenomenon happens in glasses. 
This discussion has interesting consequences if one studies the
deformation of mesoscopic samples on very small scales.
We expect that the stress $\langle \sigma \rangle$ and
shear-modulus $\mu$ of a {\it single} realization of the random walls
will be non-zero even after the thermal averaging but fluctuate along the
$\gamma$-axis (See Fig.~\ref{fig-walls-stress-strain-rem}). Only if
one performs an  average along the $\gamma$-axis will one recover the
zero average value.
Physically the breakdown of the commutation of the two limits means
that elasticity theory fails. Thus elasticity and plasticity must
emerge simultaneously in solids. 
We will argue that the plastic
events can be viewed as changes of the relevant metastable states  when
one varies $\gamma$.  (see Fig.~\ref{fig-walls-stress-strain-rem}).

{\bf Shear on a cloned system -} Let us now analyze the static response of glasses to shear.
Taking the view of the RFOT, we suppose that there exist exponentially
many metastable
states $\alpha=1,2,\ldots$  with free-energies per particle $f_{\alpha}$.

Our strategy is to consider a cloned system: $m$ replicas ($a=1,2,\ldots,m$) 
are forced to stay in the same metastable state,  and we examine how the
system responds to a generalized shear such that each replica $a$ is
submitted to
a different strain $\gamma_{a}$. 
The total free-energy $F_{m}$ of such a cloned system can
be formally as
\beq
F_{m}(\gamma)=F_{m}(0)+ N \sum_{a=1}^{m} \langle \sigma_{a} \rangle \gamma_a
+\frac{N}{2} \sum_{a,b=1}^{m} \mu_{ab} \gamma_{a} \gamma_{b} +
O(\gamma^{3}) \ .
\eeq
By clone symmetry the generalized shear-modulus $\mu_{ab}$ can be
written as
\beq
\mu_{ab} =\hat{\mu} \delta_{a,b} + \tilde{\mu}
\eeq
with 
$\hat{\mu}=\langle b \rangle -N \beta \sum_{\alpha}[\langle \sigma^{2} \rangle_{\alpha} 
-\langle \sigma \rangle_{\alpha}^{2}]P_{\alpha}$ and 
$\tilde{\mu}= - N \beta
\left[\sum_{\alpha}\langle \sigma \rangle _{\alpha}^{2}P_{\alpha}-\left(\sum_{\alpha}\langle
\sigma \rangle_{\alpha}P_{\alpha}\right)^{2} \right]$
where $P_{\alpha}=e^{-m\beta (N f_{\alpha}-F_{m}(0))}$ is the thermal weight
of the $\alpha$-th metastable state at temperature $T/m$ and 
$\langle \ldots \rangle_{\alpha}$ stands for a thermal average within the $\alpha$-th
metastable state. 
$\hat{\mu}$ can be naturally
interpreted as {\it intra-state shear-modulus} and $\tilde{\mu}$ as the
negative correction due to {\it inter-state} thermal fluctuations. 
It is easy to see that the physical shear-modulus $\mu$ of a single system
at temperature $T/m$ can be obtained as
$
\mu= \sum_{b=1}^{m}\mu_{ab}=\hat{\mu}+m \tilde{\mu}$.

Physically the intra-state shear-modulus $\hat{\mu}$ should be
interpreted as the plateau modulus measured in linear visco-elastic
measurements \cite{Weitz97}.
We expect it not to fluctuate between different metastable states. On the other hand,
$\tilde{\mu}$ which is due to the inter-state fluctuations should be different on different realizations of the random walls.
As we discussed before, the statistical translational symmetry of the
random walls requires the total modulus $\mu$ to vanish on average;
this imposes that, on average over the realizations of random walls, $\tilde \mu=-\hat \mu/m$.

Within the RFOT \cite{RFOT}, the metastable states 
disappear at a dynamical transition temperature $T_{\rm c} ( > T_{\rm K})$
predicted by the MCT \cite{MCT}. Thus the plateau modulus $\hat{\mu}$ is
positive only below $T_{\rm c}$. 

{\bf Cage expansion of the shear-modulus -} Within the cloned liquid
theory (see details in \cite{mezard-parisi-1999})  one assumes that
particles in different replicas form molecules with a certain `cage' size $A$,
which plays the role of an order parameter 
that distinguishes  the liquid phase $A=\infty$ from the glass phase $A < \infty$.
The system is considered as a liquid at an effective
temperature $T^{*}=T/m^{*}$ where  $m^{*}=m^{*}(T)$ is determined for
each temperature $T$ by the stationnarity condition of the free
energy. 
One finds that $m^*<1$ when $T<T_{\rm K}$,
 where $T_{\rm K}$ is the Kauzmann temperature\cite{Kauzmann}, while
 $m^*$ sticks to the value $1$ for larger temperatures. In many cases
 the behaviour of $m^*$ is well approximated by
$m^{*}\simeq T/T_{\rm K}$ 
\cite{mezard-parisi-1999,coluzzi-mezard-parisi-verrocchio-1999}.
It is convenient to label the molecules as $i=1,2,\ldots,N$ and to
write the position of a particle as  ${\bf r}^{a}_{i}={\bf r}_{i}+{\bf
u}^{a}_{i}$ where ${\bf r}_{i}$ is the center of mass position of the molecule 
and ${\bf u}^{a}_{i}$ describes the displacement of the particle in
replica $a$  within the molecule ($\sum_a u_i^a=0$).
The cage size  $A \equiv (N m(m-1))^{-1}\sum_{i}\sum_{a < b}\langle (u_{i}^{a}-u_{i}^{b})^{2}\rangle$ 
is assumed to be small enough to allow a small cage expansion.
Then the fluctuations within the cage is characterized by
$
\langle u_{i}^{a}u_{j}^{b} \rangle_{\rm cage} =-2(1-m\delta_{ab})\frac{A}{m}\delta_{ij}$,
where $u$ is a component of ${\bf u}$.
If the cage size $A$ changes discontinuously from a finite value to $\infty$ at $T_{\rm c}$, as predicted by the MCT \cite{MCT}, the cage expansion can work in principle right up to $T_{\rm c}$ from below \cite{parisi-zamponi}.

Using the above prescription, we have computed the shear-modulus up to 1st
order in the cage expansion\cite{us}. 
The intra-state  (or plateau)  modulus is obtained as
\begin{eqnarray}
\hat{\mu}=\langle b \rangle_{*} -J_{1}+(J_{2}+J_{3})(1-m) 
\label{eq-intra-state-shear-modulus}
\end{eqnarray}
where $\langle \ldots \rangle_{*}$ is a thermal average at temperature $T^{*}$ and,
\begin{eqnarray}
&& J_{1}=2 \frac{A}{m}\frac{1}{N}\sum_{i}\sum_{j_{1} (\neq i)}\sum_{j_{2} (\neq i)} 
\beta^{*}
\left \langle 
\nabla_{i　j_{1}} \sigma_{i　j_{1}} \cdot \nabla_{i　j_{2}} \sigma_{i　j_{2}}
\right \rangle_{*}  \nonumber \\
&& J_{2}=-2\frac{A}{m}\frac{1}{N} \sum_{i < j} 
\left \langle 
\nabla^{2}_{ij} b_{ij}
\right \rangle_{*}  \nonumber \\
&& J_{3}=2\frac{A}{m} \frac{1}{N}\sum_{i < j}\sum_{k < l}
\beta^{*}
\langle 
 b_{ij} \nabla^{2}_{kl} v(r_{kl})
\rangle_{c*}
\label{eq-J1-J2-J3}
\end{eqnarray}
where $\sigma_{ij}$ and $b_{ij}$ are the summands in \eq{eq-def-sigma-born} and
$\langle \ldots \rangle_{c*}$ stands for a connected correlation function at $T^{*}$. In the derivation of the above result, we used the fact that  the
shear-modulus is zero in the liquid  $\langle b
\rangle_{*}-\beta^{*}\langle \sigma^{2} \rangle_{c*}=0$.

The remarkable fact is that we can compute the plateau modulus at
temperatures between $T_K$ and $T_c$. From
\eq{eq-intra-state-shear-modulus}, one finds that in this regime
(where $m^*=1$), it takes
the simple and suggestive form 
$\hat{\mu}=\langle b \rangle -
J_{1}$. The physical interpretation of this result is very simple. On
time scales shorter than the $\alpha$-relaxation time, the stress-field is essentially frozen in time. 
There the only appreciable fluctuations are those associated with the $\beta$-relaxation. The term $J_{1}$ represents
the strength of stress fluctuations due to these processes.

\begin{figure}[t]
\includegraphics[width=0.25\textwidth]{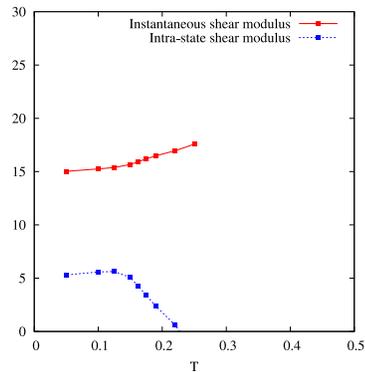}
\caption{Shear modulus of a binary soft-sphere system, computed from
  the cloned liquid approach. The bottom curve gives the plateau shear
  modulus, it is positive below  $T_c\simeq 0.22$. 
} 
\label{fig-shearmodulus}
\end{figure}

{\bf A test case: a binary soft-sphere system -} To test our scheme we
performed an explicit computation of the shear modulus of the
standard binary mixture of particles with soft-core interactions
\cite{binary-soft-sphere}. The Kauzmann temperature 
of this system is $T_{\rm  K}\simeq 0.14$
\cite{coluzzi-mezard-parisi-verrocchio-1999} while the dynamical (MCT) transition
temperature is $T_{\rm c}
\simeq 0.22$ \cite{roux-barrat-hansen}. 
In Fig.~\ref{fig-shearmodulus} we show the result for the case of
density $\rho=1$. To evaluate $m^{*}$ and the radial distribution function 
$g_*(r)$ at $T^*$, we performed the cloned liquid computation using
the binary HNC approximation \cite{coluzzi-mezard-parisi-verrocchio-1999}.

The evaluation of various terms in \eq{eq-intra-state-shear-modulus} is done as follows.
The Born term and $J_{2}$  involve only 2-point functions so that
they can be evaluated easily using $g_{*}(r)$.
We evaluate $J_{1}$ by
$
J_{1}=-2\frac{A}{m} \int d^{3}r g_{*}(r) \beta^{*}
|\nabla \sigma|^{2} (r)
 -2\frac{A}{m}\int d^{3}r_{1}d^{3}r_{2}
g({\bf r}_{1})g({\bf r}_{2})g({\bf r}_{12})
\beta^{*}\nabla \sigma ({\bf r_{1}}) \cdot \nabla \sigma ( {\bf r_{2}})
$
where we made a chain approximation in order to approximate the
three-point correlation function by a product of two-point terms, an
approximation which is reasonable at high densities. 
The evaluation of $J_{3}$ involves a
connected 4-point function
which is expected to be smaller than the
other terms and we have neglected it at present.

As shown in Fig~\ref{fig-shearmodulus}, the plateau-modulus $\hat{\mu}$
strongly depends on the temperature. Remarkably, it continuously crosses $0$ at
a temperature very close to $T_{c}$ determined by direct numerical
simulations \cite{roux-barrat-hansen}. 
Furthermore the cage size $A$ is found to be still very small
($10^{-2}$) at this temperature, which justifies our use of  the first
order small cage expansion. Our result of a plateau shear modulus
emerging continuously below $T_{\rm c}$ disagrees with the conventional
MCT \cite{MCT}, which predicts that the shear-modulus jumps
discontinuously to a finite value at $T_{\rm c}$ from the liquid side.
Our result means that the density field is frozen at $T_{c}$  as
the MCT predicted, but the system is just marginally stable there \cite{Otsuki-Sasa}, a picture which is consistent with the energy
landscape picture of the RFOT \cite{Franz-Parisi-1997,Kurchan-Laloux,Franz-Parisi-2000,Grigera-2002,berthier}.  
For the visco-elastic measurements, this continuous transition suggests
 a power law behaviour $G'(\omega), G"(\omega) \propto \omega^{\lambda}$. 

{\bf Intermittency of static shear response - } 
Our results have a natural interpretation within
a mean-field picture. They suggest the following
``intermittent'' nature of static shear response below $T_{\rm K}$ at
{\it mesoscopic scales} such that the system size $N$ is {\it large but finite}. 
Within mean field, the parameter $m=m^{*}(T)\simeq T/T_{\rm K}$ obtained in the cloned
liquid approach
 is naturally interpreted as the Parisi parameter of the 1 step replica symmetry breaking (1RSB) Ansatz for the glass phase. 

As shown in Fig.~\ref{fig-walls-stress-strain-rem}, the interpretation
at the mean field level is that of a free-energy landscape
 $F(\gamma)$ which  may be viewed as sequence of parabola with curvature
 $\hat{\mu}$ (plateau-modulus) along the $\gamma$-axis, matching with
 each other at yield points \cite{Bouchaud-Mezard-1997}.
Below $T_{\rm K}$, the static response to shear is dominated 
by intra-state response with occasional inter-state response 
when passing the yield points.

This picture is analogous to the mesoscopic response 
in mean-field spin-glass models \cite{yoshino-rizzo}. 
 Here $\gamma$ plays the role of the external magnetic field
 $h$ in  spin glasses, which exhibit step-wise increase
 of magnetization $m(h)$ along $h$-axis. The drops
 of the stress passing the yield points corresponds to steps 
of the magnetization.
At a given $\gamma$, each metastable state has a random free-energy
$f_{\alpha}$ and a random stress $\langle \sigma \rangle_{\alpha}$ so
that the increase of $\gamma$ induces  level crossings between
low-lying states.

The distribution of the stress may be modeled by a Gaussian 
distribution with zero average and variance $\Delta/\sqrt{N}$.
From the correspondence with the spin-glass problem \cite{yoshino-rizzo}
we expect  the typical spacing between the yield points to scale as
$\delta \gamma_{\rm s} \sim T_{\rm K}/(\Delta \sqrt{N})$  and the width
of thermal rounding of the yield points to  scale as  $\delta \gamma_{\rm w}
 \sim T/(\Delta \sqrt{N})$.  Here the parameter $\Delta$ is fixed as
 $\beta \Delta^{2}=\hat{\mu}/m$ in order to satisfy the condition that
 the total shear-modulus, including the inter-state shear-modulus, becomes zero on average. 
At low temperatures, if we choose a value of  $\gamma$ randomly,
most of the time we will observe the plateau modulus $\hat{\mu}$ which
is positive, and occasionally, with
probablity $\delta \gamma_{\rm w}/\delta \gamma_{\rm s} \sim T/T_{\rm
  K}$, we will find a negative shear-modulus.

{\bf Discussion - } Our computations predict a non-zero plateau-modulus $\hat{\mu}$ 
at all  temperatures below the dynamical transition temperature
$T_{\rm c}$, including in the low temperature regime
\cite{quasi-static-zero-temperature}. 
They also give a natural way to compute this dynamical transition
temperature within the cloned liquid theory, offering an alternative
to the MCT computation.
This plateau modulus should be observable dynamically on
time scales smaller than the $\alpha$ relaxation time $\tau_\alpha$. Therefore one
expects it to be seen, on all laboratory time scales, at all
temperatures below the glass transition temperature (where the
$\alpha$ relaxation time becomes larger than $10^3$s).

The prediction of  intermittent shear response in mesoscopic samples
should also  be amenable to experimental tests. It is supposed to  take place
even at temperatures higher than $T_{\rm K}$ at the length and time scales
of the so-called mosaic states proposed by the RFOT \cite{RFOT,Mosaic}
because each mosaic is subjected to a
random pinning field provided by surounding mosaics.

{\bf Acknowledgment} 
We thank Giulio Biroli, Jean-Philippe Bouchaud, Song-Ho Chong, Silvio
Franz, Jorge Kurchan, Anael Lema\^\i tre, Kunimasa Miyazaki, Michio Otsuki and Tommaso Rizzo for useful discussions.  
This work is supported by a {\it Triangle de la physique} grant number 117.


\begin{thebibliography}{99}
\bibitem{us} Details of the derivations will be published elsewhere.
\end{thebibliography}
\end{document}